\def\Tr{{\mathrm Tr}\ }
\def\j{{\mathbf j}}

\def\be{\begin{equation} }
\def\ee{\end{equation} }
\def\eq{\ =\ }

\def\pls{\ +\ }

\def\JPCM{{\sl J. Phys. Condens Matter\ }}
\def\prb{{\sl Phys. Rev.\ }}
\def\PR{{\sl Phys. Rev.\ }}
\def\PRL{{\sl Phys. Rev. Lett.\ }}
\def\be{\begin{equation}}
\def\ee{\end{equation}}
\def\be{\begin{equation}}
\def\ee{\end{equation}}
\def\n{\noindent\ }
\documentclass[twocolumn,showpacs,superscriptaddress,preprintnumbers,amsmath,amssymb,prl]{revtex4}
\usepackage{graphicx}
\usepackage{dcolumn}
\usepackage{color}
\usepackage{bm}
 
  \def\ket{\vert \vert  \{ \emptyset \} \rangle}
  \def\ket2{\vert \vert \otimes \{ R \} \rangle}

  \def\ket{\vert \vert  \{ \emptyset \} \rangle}
  \def\ket2{\vert \vert \otimes \{ R \} \rangle}

\def\prl#1{ Phys. Rev. Lett. {\bf #1}}
\def\.#1{\mathaccent 95#1}
\def\^#1{\mathaccent 94 #1}
\def\~#1{\mathaccent "7E #1}

\def\eq{\enskip =\enskip}
\def\pls{\enskip +\enskip}

  \def\ket{\vert \vert  \{ \emptyset \} \rangle}
  \def\ket2{\vert \vert \otimes \{ R \} \rangle}

\def\prl#1{ Phys. Rev. Lett. {\bf #1}}
\def\.#1{\mathaccent 95#1}
\def\^#1{\mathaccent 94 #1}
\def\~#1{\mathaccent "7E #1}

\def\eq{\enskip =\enskip}
\def\pls{\enskip +\enskip}

  \def\ket{\vert \vert  \{ \emptyset \} \rangle}
  \def\ket2{\vert \vert \otimes \{ R \} \rangle}
  
\def\k{{\bf k}}

\def\etal{{\sl et.al.}\ }
\begin{document}

\title {Effect of Short-ranged Order on the electronic structure and optical properties of the CuZn alloy : an augmented space approach.}  
\author{Kartick Tarafder}
\affiliation{S.N. Bose National Center for Basic Sciences, JD Block, Sector III, Salt Lake City, Kolkata 700 098, India }
\author{Atisdipankar Chakrabarti \footnote{Permanent address : Ramakrishna Mission Vivekananda Centenary College, Rahara, West Bengal, India}}
\affiliation{S.N. Bose National Center for Basic Sciences, JD Block, Sector III, Salt Lake City, Kolkata 700 098, India }
\author{Kamal Krishna Saha}
\affiliation{Theory Department, Max-Planck-Institut f\"ur Mikrostrukturphysik,
Weinberg 2, D-06120 Halle (Saale), Germany}
\author{Abhijit Mookerjee}
\affiliation{S.N. Bose National Center for Basic Sciences, JD Block, Sector III, Salt Lake City, Kolkata 700 098, India }
\date{\today}
\begin{abstract}
We report here a study of the effect of short-ranged ordering on the electronic
structure and optical properties of CuZn alloys. We shall use the augmented space
recursion technique developed by us in conjunction with the tight-binding linear muffin tin orbitals basis. 
\end{abstract}
\maketitle
\section{Introduction}
\begin{figure}[h]
\vskip 0.25in \centering
\includegraphics[width=3in,height=2.5in]{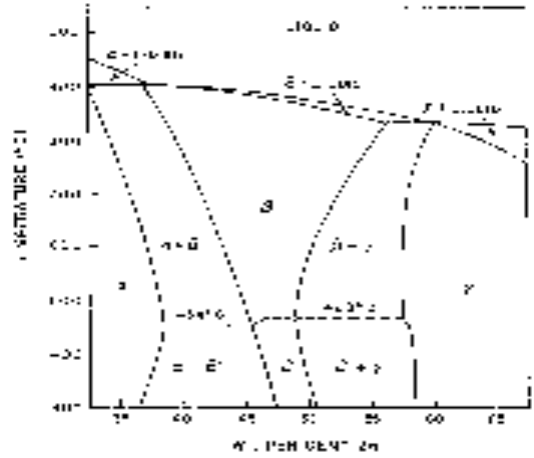}
\caption{Experimental phase diagram of CuZn alloy}
\label{pd}
\end{figure}

Binary alloys involving equal proportions of a noble metal Cu and a divalent metal Zn have a stable 
low temperature $\beta$-phase which sits immediately to the right of the pure face-centered cubic Cu phase in the alloy phase diagram \cite{jm,mas}. This phase called $\beta$ brass has a B2 cubic structure with two atoms per unit cell. At high temperatures the alloy
forms a disordered body-centered cubic structure. The disorder-order transition
takes place around 730K.
 The alloy satisfies the Hume-Rother\'y rules 
\cite{hr}  and has the same ratio of valence electrons to atoms. Jona and Marcus \cite{jm} have shown from a Density Functional Theory (DFT) based approach that
within the Local Density Approximation (LDA), it is the body-centered based B2 
which is the stable ground state. They also showed that if we include the Gradient Corrections (GGA) then we get a tetragonal ground state lower in energy by 0.1 mRy/atom.
This is in contradiction with the latest experimental data. The alloying of face-centered cubic Cu with an equal amount of Zn leads to a body-centered stable phase. Zn has only one more electron than Cu. This is an interesting phenomenon. CuZn alloys also have anomalously high elastic anisotropy. This makes the theoretical study of CuZn an interesting exercise for a proposed theoretical technique.
The phase diagram of CuZn is shown in Fig. \ref{pd}.

One of the  earliest first-principles Density functional based study of the electronic properties of CuZn
as by Bansil \cite{bansil}. The authors  had  studied the complex bands of $\alpha$-phase of CuZn using the Korringa-Kohn-Rostocker (KKR) method coupled with the coherent potential approximation (CPA) to take care of disorder. They commented on the effects of charge transfer and lattice constants on the electronic structure. They found the  electronic distribution of this alloy to be of
 a split band kind with the centers of the Cu
and Zn $d$-bands well separated from each other. Their Zn $d$-bands showed hardly any dispersion and were shown only schematically in their figures. In a later work Rowlands \cite{row} generalized the CPA to a non-local version (NL-CPA) and studied the effects of short-ranged ordering in CuZn. Their technique was based on an idea of renormalization in reciprocal space suggested by Jarrel and Krishnamurthy \cite{jar}.  

The order-disorder transition in CuZn is a classical example of a true second order transition. Several very early investigations on this alloy has been reviewed by Nix and Shockley \cite{nix} and Guttman \cite{gutt}. These investigations, of course, were rather crude, as sophisticated approaches 
to deal with disordered alloys had not really been developed at that time.  However, it was recognized that a knowledge of the short-ranged order correlations above the critical temperature should be
of considerable interest. Early neutron scattering experiments were carried out on $\beta$-brass
by Walker and Keating \cite{wk}.  The Warren-Cowley short-ranged order parameter, defined by
$\alpha(R)=1-P_{AB}(R)/x$ where x is the concentration of A, $P_{AB}(R)$ is the probability of finding an A atom at a distance  of R from a B atom,  was directly obtained from the diffuse scattering cross-section :

\[ \frac{d\sigma}{d\Omega}\ =\ x(1-x)(b_A-b_B)^2 \sum_R \alpha(R) f(K)\exp(iK\cdot R)\]

where $b_A,b_B$ are the scattering lengths of $A$ and $B$ atoms, and $f(K)=\exp(-C\vert K\vert^2)$ is the attenuation factor arising from thermal vibrations and static strains. The experimental data for
the short-ranged order parameter as a function of temperature are thus available to us.
A Ising-like model using pair interactions was studied by Walker and Chipman \cite{wc} and the
short-ranged order was theoretically obtained. However, the pair interactions were simply fitted
to the experimental values of the transition temperature $T_C$ and in that sense it was an
empirical theory. The experimental estimate of the nearest neighbour Warren-Cowley parameter
was found to be varying between -0.171 to -0.182 at around 750K. 

In a later work using the much more sophisticated {\sl locally self-consistent Green function} (LSGF)
approach based on the tight-binding linear muffin-tin orbital (TB-LMTO) technique Abrikosov \etal
 \cite{abri} studied CuZn alloys. The authors argued that earlier studies of the mixing enthalpies of CuZn using the standard coherent potential approximation approaches \cite{cpa1}-\cite{cpa5}
showed significant discrepancy with experiment. Part of the discrepancy was assumed to be
from neglect of charge transfer effects and part from the effects of short ranged ordering (SRO).
The main thrust of this technique, which was based on an earlier idea of a locally self-consistent multiple scattering (LSMS) by Wang \etal \cite{wang}, was to go beyond the CPA and 
include the effects of the immediate environment of an atom in the solid.  The LSMS gave an excellent
theoretical estimate of the ordering energy in CuZn : 3.37 MRy/atom as compared to the experimental
value of 3.5 MRy/atom.  The LSGF approach correctly predicted ordering tendency in CuZn on lowering 
temperature and combing with a Cluster Variation-Connolly Williams (CVM-CW) obtained a value of the nearest neighbour 
Warren-Cowley SRO parameter $\alpha$ = -0.15.  Subsequently Bruno \etal \cite{bruno} proposed a modification of the CPA taking into account the local field and showed that charge transfer effects can be taken into account accurately as compared to the O(N) methods just described. They applied their approach to the CuZn alloys.

One of the earlier works on the optical property of CuZn alloy was the determination of the temperature variation of spectral reflectivity by Muldawer \cite{mul}. The author attempted to explain
the color of the disordered $\beta$-brass CuZn alloy via the internal photoelectric effect \cite{jk}.
Although, the experimental data also contained the contribution from plasma oscillations, the author claimed that the spectral reflectivity helps to  explain the band picture of the alloys as a function of the inter-atomic spacing. In order to explain the optical properties, Amar \etal \cite{amar1}-\cite{amar3} studied the band structure of CuZn using the KKR method. However, they had used the virtual crystal approximation, replacing the random potential seen by the electrons by an averaged one. This is now known to be particularly inaccurate for split band alloys. 

The above discussion was necessary to bring into focus the following points~: in the study of alloys like CuZn it would be interesting to address the effects of charge transfer and short-ranged ordering. In this communication we shall address exactly these two points. We shall propose the use of the augmented space recursion (ASR) coupled with the tight-binding linear muffin-tin orbitals basis (TB-LMTO) \cite{asr}
to study the effects of short-ranged ordering on both the electronic structure and the optical properties of $\beta$-CuZn alloy at 50-50 composition.
We should like to stress here that the TB-LMTO-ASR addresses precisely these
effects with accuracy : the Density functional self-consistent TB-LMTO takes care of the charge transfer, while the local environmental effects which are essential for the description of SRO are dealt with by the ASR. 
The TB-LMTO-ASR and its advantages  has been extensively discussed earlier by Mookerjee \cite{tf} and in a series of articles  \cite{asr,Am,KG,asr2,Am3,Am4}. We would like to refer the interested readers to these for details. 

\section{Spectral functions, Complex bands and Density of States for 50-50 CuZn}

In this section we shall introduce the salient features of the ASR  which will be required by us in our
subsequent discussions.

We shall start from a first principle TB-LMTO set of 
orbitals  \cite{Ander1,Ander2} in the most-localized
$\alpha$ representation. This is necessary, because the subsequent
recursion requires a sparse representation of the Hamiltonian. In this 
representation, the second order alloy Hamiltonian is given by,

\[{\mathbf H}^{(2)} = {\mathbf E}_\nu + {\mathbf h} - \mathbf{hoh}  \]
where,
\begin{eqnarray}
{\mathbf h} \!\! &=& \!\!\!  \sum_{R} \left({\bf C}_{R} - {\mathbf E}_{\nu R}\right) {\cal P}_{R} 
 + \sum_{R}\sum_{R'}{\mathbf \Delta}_{R}^{1/2} \ {\mathbf S}_{RR'} \ {\mathbf \Delta}_{R'}^{1/2} \ {\cal T}_{RR'}\nonumber\\ 
\phantom{x}\nonumber\\
{\mathbf o} \!\! &=& \!\!\! \sum_R \mathbf{o}_R \ {\cal P}_R 
\end{eqnarray}

\noindent {\bf C}$_R$, {\bf E}$_{\nu R}$ , ${\mathbf\Delta}_R$  and ${\mathbf o}_R$ are diagonal matrices in angular momentum space, 
 and {\bf S}$_{RR'}$ is the structure matrix. These matrices are  of rank $L_{max}$.
 ${\cal P}_{R}\ =\ \vert R\rangle\langle R\vert$ and ${\cal T}_{RR'}\ =\ 
\vert R\rangle\langle R'\vert$ are projection and transfer operators in the Hilbert
space ${\cal H}$ spanned by the tight-binding basis $\{|R\rangle\}$. Here, $R$ refers 
to the position of atoms in the solid  and $L$ is a composite label $\{\ell,m,m_s\}$ for 
the angular momentum quantum numbers. ${\bf C}$, ${\bf \Delta}$ and ${\bf o}$ are the potential 
parameters of the atoms which sit on the lattice sites ; ${\bf o}^{-1}$ has dimension of energy and ${\bf E}_\nu$'s are 
the reference energies about which the muffin-tin orbitals are linearized.

For a disordered binary alloy we may write~: 

\begin{eqnarray}
{ C}_{RL} & = &  C_L^A\  n_R + C_L^B\ \left( 1-n_R \right) \nonumber \\
\Delta_{RL}^{1/2} & = & \left( \Delta_L^A \right)^{1/2} n_R 
+ \left(\Delta_L^B \right)^{1/2}\left( 1-n_R \right)  \nonumber \\
{o}_{RL} & = &  o_L^A\  n_R + o_L^B\ \left( 1-n_R \right) 
\end{eqnarray}

Here $\{n_R\}$ are the random site-occupation variables which take values 1 and 0 
 depending upon whether the muffin-tin labelled by $R$ is occupied by an $A$ or
a $B$-type of atom. The atom sitting at $\{R\}$ can either be of type $A \ (n_R=1)$
with probability $x$ or $B \ (n_R=0)$ with probability $y$, where $x$ and $y$
are the concentrations of the components $A$ and $B$ in the binary alloy.
The augmented space
formalism (ASF) now introduces the space of configurations of the set of binary 
random variables  $\{n_R\}$ : $\Phi$. 
In the absence of short-ranged order, each random variable  $n_R$ has associated with it an operator  ${\bf M}_R$
whose spectral density is its probability density :

\begin{eqnarray}
p(n_R) &=& x \delta(n_R-1) + y\delta(n_R) \nonumber \\ 
&=& -\frac{1}{\pi} \lim_{\delta\rightarrow 0} \ \mbox{Im} 
\langle \uparrow_R|\left((n_R+i\delta) {\bf I}-{\bf M}_R\right)^{-1}|\uparrow_R\rangle. \nonumber\\ 
\end{eqnarray}

\noindent ${\bf M}_R$ is an operator whose eigenvalues $1, 0$ correspond to the observed values of $n_R$ and
whose corresponding eigenvectors $\{|1_R\rangle, |0_R\rangle\}$ span a configuration space
$\phi_R$ of rank 2.  We may change the basis to  $\{|\uparrow_R\rangle,|\downarrow_R\rangle\}$ 
\begin{eqnarray*}
 \vert\uparrow_R\rangle = \left\{ \sqrt{x}\vert 0_R\rangle+\sqrt{y}\vert 1_R\rangle\right\} \\
 \vert\downarrow_R\rangle = \left\{ \sqrt{y}\vert 0_R\rangle-\sqrt{x}\vert 1_R\rangle\right\} 
\end{eqnarray*}
and in the new basis the operator ${\bf M}_R$ corresponding to $n_R$ is~: 

\begin{eqnarray*}
 {\bf M}_R &= & x{\cal P}^\uparrow_R + y{\cal P}^\downarrow_R + \sqrt{xy} \ {\cal T}^{\uparrow\downarrow}_R\nonumber\\
& = & x {\cal I}+ b(x) {\cal P}^\downarrow_R + f(x) {\cal T}^{\uparrow\downarrow}_R 
\end{eqnarray*}

where ${\cal P}^{\uparrow}_R = \vert {\uparrow}_R\rangle\langle {\uparrow}_R\vert$, 
 ${\cal P}^{\downarrow}_R = \vert {\downarrow}_R\rangle\langle {\downarrow}_R\vert$  and 
 ${\cal T}^{\uparrow\downarrow}_R = \vert {\uparrow}_R\rangle\langle {\downarrow}_R\vert +   
 \vert {\downarrow}_R\rangle\langle {\uparrow}_R\vert$ are the projection and transfer operators in the configuration space 
$\phi_R$ spanned by the two basis vectors. 
$b(x)\ =\ (y-x)$ and $f(x)\ =\ \sqrt{xy}$.
\vskip 0.2cm
 The full configuration space $\Phi$ = $\prod^\otimes_R\ \phi_R$ is then spanned by vectors of the form $\vert\uparrow\uparrow\downarrow\uparrow\downarrow\ldots\rangle$.
These configurations may be labelled by the sequence of sites $\{{\cal C}\}$ 
 at which we have a $\downarrow$. For example, for the state just quoted  $\{{\cal C}\}$
= $\vert\{3,5,\ldots\}\rangle$. This sequence is called the {\sl cardinality sequence}.
If we define the configuration $\vert\uparrow\uparrow\ldots\uparrow\ldots\rangle$ as the 
 {\sl reference} configuration, then the {\sl cardinality sequence} of the {\sl reference} configuration is the null sequence 
$\{\emptyset\}$.

In the full augmented space the operator corresponding to $n_R$ is :

\begin{eqnarray}
\widetilde{\bf M}_R 
     =  x \ {\cal I}\otimes {\cal I}\ldots + b(x)\ {\cal P}^\downarrow_R\otimes {\cal I}\ldots 
  +  f(x)\ {\cal T}^{\uparrow\downarrow}_R\otimes{\cal I}\ldots\nonumber\\
\label{asr1}
 \end{eqnarray}

The augmented space theorem \cite{Am} states that

\be 
\ll A(\{ n_{R}\}) \gg \eq < \{\emptyset\}\vert \widetilde{{\mathbf A}}(\{{\mathbf M}_R\})\vert \{\emptyset\}>,
\label{eq5}
\ee

\n where,

\[ \widetilde{\mathbf A}(\{{\widetilde{\bf M}_R}\}) \eq \int \ldots \int A(\{\lambda_{R}\})\ \prod d{\mathbf P}(\lambda_{R}).\]

\n {\bf P}($\lambda_{R}$) is the spectral density of the self-adjoint operator $\widetilde{{\bf M}}_{R}$. 
Applying (\ref{eq5}) to the Green function we get~:
\begin{equation}
\ll {\bf G}(\k,z)\gg \ = \ \langle \k\otimes\{\emptyset\} |{(z{\bf\widetilde I} 
- {\bf\widetilde{H}}^{(2)})}^{-1} |\k\otimes\{\emptyset\} \rangle.
\label{eq6}
\end{equation}

where {\bf G} and {\bf H}$^{(2)}$ are operators which are matrices in angular momentum space, and 
the augmented reciprocal space basis $|\k,L\otimes\{\emptyset\} \rangle$ has the form

\[ (1/\sqrt{N})\sum_R \mbox{exp}(-i\k\cdot R)|R, L\otimes \{\emptyset\}\rangle. \]

The augmented space Hamiltonian ${\bf\widetilde H}^{(2)}$ is constructed from the TB-LMTO
Hamiltonian ${\bf H}^{(2)}$ by replacing each random variable $n_R$ by  the operators $\widetilde{\bf M}_R$.
It is an operator in the augmented space 
$\Psi$ = ${\cal H} \otimes \Phi$. The ASF maps 
a disordered Hamiltonian described in a Hilbert space ${\cal H}$ onto an ordered Hamiltonian 
in an enlarged space $\Psi$, where the space $\Psi$ is constructed as the outer product of the
space ${\cal H}$ and configuration space $\Phi$ of the random variables of the disordered 
Hamiltonian. The  configuration space $\Phi$ is of rank 2$^{N}$ if there are $N$ muffin-tin 
spheres in the system. Another way of looking at ${\bf\widetilde H}^{(2)}$ is to note
that it is the {\sl collection} of all possible Hamiltonians for all possible
configurations of the system.

Combining (2) and (4) we get for any of the operators {\bf V} diagonal in real  
and angular momentum spaces :

\begin{widetext}
\begin{eqnarray}
\widetilde{\bf V}= \sum_R\left\{\rule{0mm}{5mm}{\bf A}({\bf V})\ {\cal P}_R\otimes{\cal I}\otimes{\cal I}\ldots 
+ {\bf B}({\bf V})\ {\cal P}_R\otimes{\cal P}^\downarrow_R \otimes{\cal I}\ldots
\pls {\bf F}({\bf V})\ {\cal P}_R\otimes{\cal T}^{\uparrow\downarrow}_{R}\otimes{\cal I}\ldots 
\right\}\ =\ \tilde{\mathbf A}+\widetilde{\mathbf B}+\widetilde{\mathbf F}.
\label{eq24}
\end{eqnarray}
\end{widetext}
where for any of the diagonal (in real and angular momentum spaces) operator {\bf V} : 
\begin{eqnarray}
{\mathbf A}({\mathbf V}) = & (x\ V^A_L + y\ V^B_L)\  \delta_{LL'}\nonumber\\
{\mathbf B}({\mathbf V}) = &b(x)(V^A_L-V^B_L)\ \delta_{LL'}\nonumber\\
{\mathbf F}({\mathbf V}) = & f(x)(V^A_L-V^B_L)\ \delta_{LL'}\nonumber\\ 
\label{def}
\end{eqnarray}

In case there is no off-diagonal disorder due to local lattice distortion because of size mismatch :

\[ {\bf\widetilde S} = \sum_{R}\sum_{R'} {\bf A}({\bf \Delta}_R^{-1})^{-1/2}\ {\bf S}_{RR'} \ {\bf A}({\bf \Delta}_{R'}^{-1})^{-1/2} \enskip
{\cal T}_{RR'}\otimes {\cal I}\otimes {\cal I}\ldots. \]

This equation is now exactly in the form in which the recursion method
\cite{vol35}  may be applied. At this point
we note that the above expression for the averaged $G_{LL}(\k,z)$ is {\sl exact}.

The recursion method addresses inversions of infinite matrices. Once a sparse representation
of an operator in Hilbert space, ${\bf\widetilde H}^{(2)}$, is known in a countable basis, the recursion method
obtains an alternative basis in which the operator becomes tridiagonal. This basis and the 
representations of the operator in it are found recursively through a three-term recurrence relation~:
\begin{equation}
|u_{n+1}\} = {\bf\widetilde H}^{(2)} |u_n\} - \alpha_n(\k) |u_n\} - \beta_n^2(\k) |u_{n-1}\}.
\end{equation}

with the initial choice $|u_1\}=|{\mathbf k}L\rangle\otimes|\{\emptyset\}\rangle$ and $\beta_1^2=~1$. The recursion 
coefficients $\alpha_n$ and $\beta_n$ are obtained by imposing the ortho-normalizability 
condition of the new basis set as~:

\begin{eqnarray*}
&&\alpha_n(\k) = \frac{\{n|{\bf\widetilde H}^{(2)}|n\}}{\{n|n\}} \phantom{x} ; \phantom{xx} \beta_{n-1}^2(\k) = \frac{\{n-1|{\bf\widetilde H}^{(2)}|n\}}{\{n|n\}} \phantom{x} \\
&& \mbox{and also} \phantom{xx} \{m|{\bf\widetilde H}^{(2)}|n\} = 0 \mbox{  for } m\not= n, n\pm 1
\end{eqnarray*}

To obtain the spectral function we first write the configuration averaged $L$-projected Green
functions as continued fractions~:

\begin{widetext}
\[ \ll G_{LL}(\k,z) \gg \ = \ \frac{{\beta_{1L}^2}}
        {\displaystyle z-\alpha_{1L}(\k)-\frac{\beta^2_{2L}(\k)}
        {\displaystyle z-\alpha_{2L}(\k)-\frac{\beta^2_{3L}(\k)}
        {\displaystyle \frac{\ddots}
        {\displaystyle z-\alpha_{NL}(\k)-\mathbf \Gamma_L(\k,z)}}}}. \]
\end{widetext}

where $\mathbf \Gamma_L(\k,z)$ is the asymptotic part of the continued fraction.  
The approximation involved has to do with the termination of this continued fraction. The coefficients 
are calculated exactly up to a finite number of steps $\{\alpha_n,\beta_n\}$ for $n < N$ and the asymptotic
part of the continued fraction is obtained from the initial set of coefficients using the idea of Beer and
Pettifor terminator \cite{bp}. Haydock and coworkers \cite{kn:hay} have carried out extensive studies of 
the errors involved and precise estimates are available in the literature. Haydock \cite{kn:hay2} has shown
that if we carry out recursion exactly up to $N$ steps, the resulting continued fraction maintains the first 
$2N$ moments of the exact result.

It is important to note that the operators ${\bf\widetilde A}, {\bf\widetilde B}, {\bf\widetilde F}$ are all projection operators
in real space ({\sl i.e} unit operators in {\bf k-} space) and acts on an augmented space basis only to 
change the configuration part ({\sl i.e.} the cardinality sequence $\{{\cal C} \}$ ).

\begin{eqnarray*}
{\bf\widetilde A} \vert\vert \{ {\cal C} \} \rangle &=& A_{1} \vert\vert \{ {\cal C} \}\rangle, \\
{\bf\widetilde B} \vert\vert \{ {\cal C} \} \rangle &=& A_{2} \vert\vert \{ {\cal C} \} \rangle \ 
\delta( R \in \{ {\cal C} \}),\\
{\bf\widetilde F} \vert\vert \{ {\cal C} \} \rangle &=& A_{3} \vert\vert \{ {\cal C} \pm R \} \rangle.
\end{eqnarray*}

The coefficients $A_{1} - A_{3}$ can been expressed from equation (\ref{def}). 
 The remaining operator ${\bf\widetilde S}$ is 
diagonal in {\bf k-} space and acts on an augmented space only to change the configuration part~:

\begin{equation}
{\bf\widetilde S} \vert\vert \{ {\cal C} \} \rangle = \sum_{\chi} \exp{(-\imath {\bf k}.\chi) } \vert\vert \{ {\cal C} - \chi \} \rangle. \nonumber
\end{equation}

\begin{figure*}
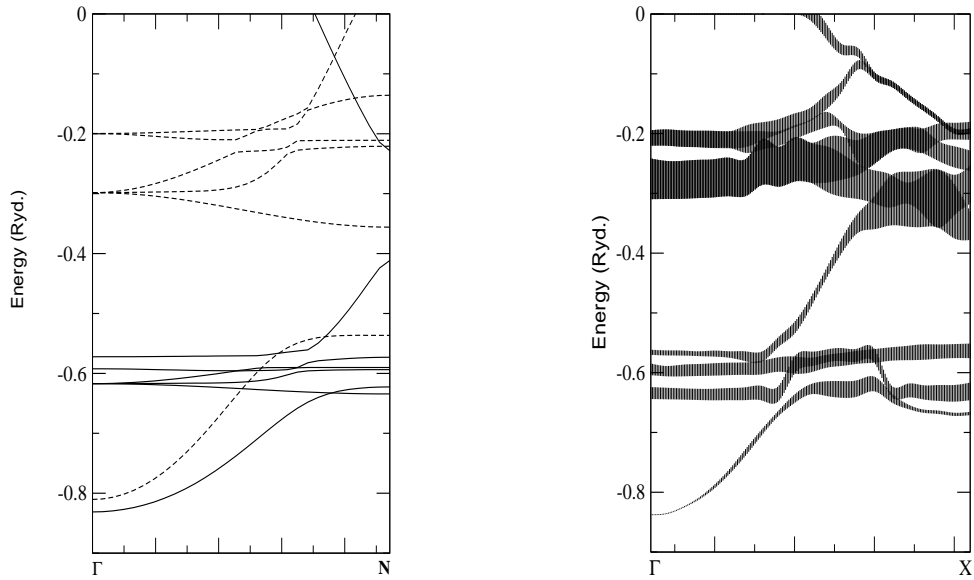

\centering
\includegraphics[width=2in,height=3in]{fig2.eps} \hskip 1in
\includegraphics[width=2in,height=3in]{fig3.eps}
\caption{(left) Bands for pure Cu and Zn in bcc lattices with the same
lattice parameter as the 50-50 CuZn alloy. The dashed line are for Cu and the full lines for Zn. (right) Complex bands for the
50-50 CuZn alloy.}
\label{fig1}
\end{figure*}

Here $\chi$s are the near neighbour vectors. The operation of the effective Hamiltonian 
is thus entirely in the configuration space and the calculation does not involve the space 
${\cal H}$ at all. This is an enormous simplification over the standard augmented
space recursion \cite{asr,asr2,Am3,Am4}, where the entire reduced real
space part as well as the configuration part were involved in the recursion process. Earlier 
we had to resort to symmetry reduction of this enormous space in order to make the recursion 
tractable. Here the rank of only the configuration space is much smaller and we may
further reduce it by using the local symmetries of the configuration space, as described 
in our earlier letter \cite{asr}. However, this advantage is offset by the fact that the
effective Hamiltonian is energy dependent. This means that to obtain the  Green functions 
we have to carry out the recursion for each energy point. This process is simplified by 
carrying out recursion over a suitably chosen set of {\sl seed energies} and interpolating 
the values of the coefficients across the band.

The self-energy which arises because of scattering by the random potential fluctuations is
of the form :

\[ \Sigma_L(\k,z) \ = \ {\frac{\beta^2_{2L}(\k)}
        {\displaystyle z-\alpha_{2L}(\k)-\frac{\beta^2_{3L}(\k)}
        {\displaystyle \frac{\ddots}
        {\displaystyle z-\alpha_{NL}(\k)-\mathbf \Gamma_L(\k,z)}}}}.         \]

So the continued fraction can be written in the form $1/{(z-\tilde E_L(\k)-\Sigma_L(\k,E))}$, 
where $\tilde E_L(\k)=\alpha_{1L}(\k)$. 

The average spectral function $\ll A_\k(E) \gg$ is related to the averaged Green function in 
reciprocal space as :

\[ \ll A_\k(E) \gg \ = \sum_L \ll A_{\k L}(E) \gg ,\]
where
\[ \ll A_{\k L}(E) \gg \ = -\frac{1}{\pi} \lim_{\delta \rightarrow 0+} \{\mbox{Im} \ll G_{LL}(\k,E-i\delta)\gg\}. \]

\begin{figure}[b]
\vskip 0.3in
\centering
\includegraphics[width=2.5in,height=2.5in]{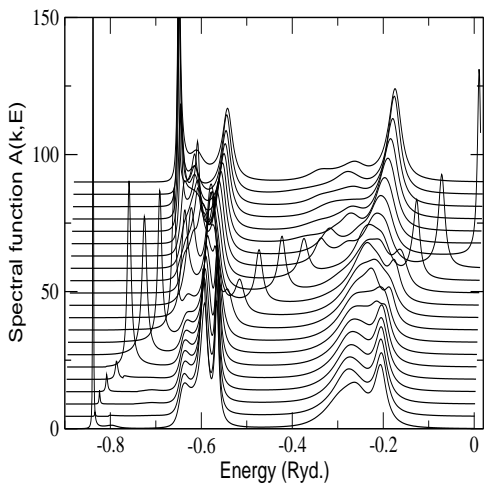}
\caption{Spectral functions for the CuZn alloy for {\bf k}-vectors along the $\Gamma$ to $X$ direction in the Brillouin zone.}
\label{fig2}
\end{figure}

To obtain the complex bands for the alloy we fix a value for $\k$ and solve for~:
\[ z-\tilde E_L(\k)-\Sigma_L(\k,E) = 0. \]
\begin{figure*}
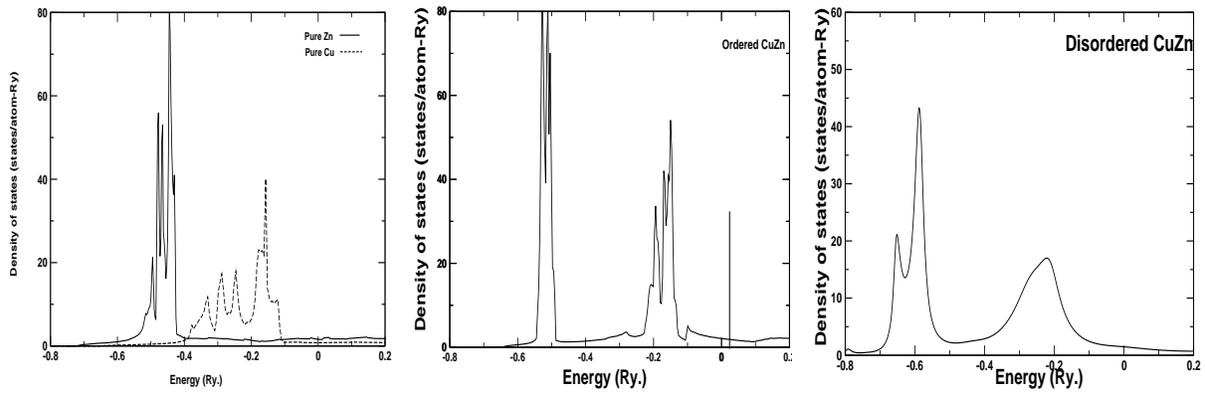

\centering
\includegraphics[width=2in,height=2in]{fig5.eps}\hskip 0.3cm
\includegraphics[width=2in,height=2in]{fig6.eps}\hskip 0.3cm
\includegraphics[width=2in,height=2in]{fig7.eps}
\caption{(left) Density of states of pure Zn (solid lines) and Cu (dashed lines)in the same bcc lattice as the 50-50 CuZn alloy. (centre) Density of states
for ordered B2 50-50 CuZn alloy. (right) Density of states for the disordered
bcc 50-50 CuZn alloy.}
\label{fig3}
\end{figure*}

The real part of the roots will give the position of the bands, while the imaginary part of
roots will be proportional to the lifetime. Since the alloy is random, the bands always have
finite lifetimes and are fuzzy.  

We have used this reciprocal space ASR to obtain the  complex bands and spectral functions for the CuZn alloy. This is shown in  Figs. (\ref{fig1}) - (\ref{fig2}).
It should be noted that we have carried out a fully LDA self-consistent calculation using the TB-LMTO-ASR developed by us \cite{adc} to obtain the potential parameters. It takes care of the charge transfer effects. For the  Madelung energy part of the alloy calculation, we have chosen the approach of Ruban and Skriver  \cite{skriver}. 

The two panels of Fig. (\ref{fig1}) compare the band structures of pure Cu and pure Zn metals in the same bcc lattice as the 50-50 alloy. We note that the $s$-like bands of Cu and Zn stretch from -0.8 Ry.,while the $d$-like states of Zn and Cu, whose degeneracies are lifted by the cubic symmetry of the bcc lattice are
more localized and reside in the neighbourhood of -0.6 Ry. and between -0.3 to -0.2 Ry. respectively. The complex bands of the solid clearly reflect the same
band structure. However, the bands are slightly shifted and broadened because of the disorder scattering of Bloch states in the disordered alloy. The scattering lifetimes are maximum for the Cu $d$-like bands, less for the Zn $d$-like bands and minimum for the lower $s$-like bands. This is expected, since the delocalized $s$-like states straddle large volumes of the lattice space and are therefore less sensitive to the local configuration fluctuations of the substitutional alloy. 
\begin{figure*}
\centering \includegraphics[width=4.5in,height=1.3in]{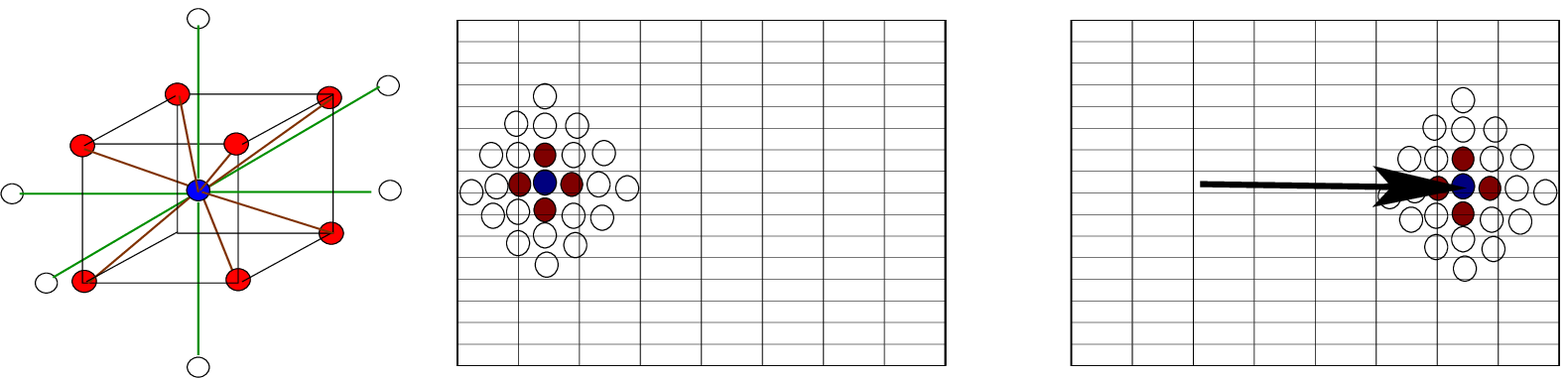}
\caption{(Color Online) (left) The blue atom represents the site labelled 0, where the local density
of states will be calculated. The red atoms are its eight nearest neighbours on the bcc lattice. The white atoms are more distant neighbours (right)\ Schematic diagram showing lattice translational symmetry of the
nearest neighbour cluster SRO approach.}
 \label{figa}
\end{figure*} 

The same is reflected in the spectral functions, shown here also along the
$\Gamma$-$X$ direction in the Brillouin zone. Sharp peaks stretching from -0.8 Ry., groups of wider peaks around -0.6 Ry. with less dispersion, characteristic of the more localized $d$-like states and groups of much wider peaks straddling
-0.3--0.2 Ry. also with less dispersion.  The spectral functions play an important role in response functions related to photoemmission  and optical conductivity \cite{km1}. Our complex bands agree remarkably well with the Fig.3 of
Bansil and Ehrenreich \cite{bansil}. These authors of course did not show
the dispersion of the Zn $d$-bands, but as in their work, the Cu bands show
greater disorder induced lifetimes than the lower energy Zn $d$-bands.

 We may use the generalized tetrahedron method to pass from the reciprocal space spectral functions to the real space density of states \cite{kspace}. Alternatively, we may also carry out real-space ASR to obtain the density of states directly.

Fig. (\ref{fig3}) shows the densities of states for the pure Zn (solid lines) and Cu (dashed lines) in the same bcc lattice as the alloy and compares this with 
the ordered B2 and disordered bcc 50-50 CuZn alloy. We first note that in the ordered B2 alloy there is a considerable narrowing of the Zn  well as the Cu $d$-like bands. The feature around -0.3 Ry. is suppressed in the ordered alloy. In the disordered alloy on the other hand, although disorder scattering introduces
life-time effects which washes out the sharp structures in the ordered systems,
the resemblance to the pure metals is evident. As seen in the complex bands, the life-time effects in the Cu $d$-like part is prominent. If we interpret the left most figure as that due to completely segregated Cu-Zn and the middle one as the completely ordered one, then the disordered alloy lies between the two. In the next section, introducing short-ranged ordering effects on top of the fully disordered alloy, we shall study how to bridge between the two states.

\section{Short-Ranged Ordering in the alloys}
Attempts at developing generalizations of the coherent potential approximation (CPA) to include effects
of short-ranged order (SRO) have been many,  spread over the last several decades. 
The CPA being a single site mean-field approximation cannot take into account SRO as it spans at least a nearest neighbour cluster on the lattice. The early attempts at
 cluster-generalizations of the CPA were beset with difficulties of violation of the analytic properties
of the approximated configuration averaged Green function.  Tsukada's \cite{mcpa} idea of introducing a 
super-cell the size of the cluster in an effective medium suffered from the problem of broken translational symmetry within the cluster even when the disorder was homogeneous. This same problem beset the CCPA's proposed by Kumar \etal \cite{ksm} based on the augmented space theorem introduced by Mookerjee \cite{Am}. The embedded cluster
method of Gonis \etal \cite{gonis} immersed a cluster in a CPA medium which lacked the full self-consistency
with it.  The first, translationally symmetric cluster approximations which preserved the analytic properties
of the approximate Green functions were all based on the augmented space theorem of Mookerjee \cite{Am}. They
included the travelling cluster approximation (TCA) of Kaplan and Gray \cite{KG} and Mills and Ratnavaraksa \cite{tca} and the CCPA proposed by Razee \etal \cite{razee}. The problem with these approaches was that they became intractable as the size of the cluster considered was increased much beyond two sites. Mookerjee and Prasad \cite{mp}
generalized the augmented space theorem to include correlated disorder. However, since they then went on to apply it in the CCPA approximation, they could not go beyond the two-site cluster and that too to model systems alone.
The breakthrough came with the augmented space recursion (ASR) approach proposed by Saha \etal \cite{sdm1}-\cite{sdm2}. The method was a departure from the mean-field approaches which always began by embedding a cluster
in an effective medium which was then obtained self-consistently. Here the Green function was expanded in a continued fraction whose asymptotic part was obtained from its initial steps through an ingenious {\sl termination}
procedure \cite{vol35}. In this method the effect at a site of quite a large environment around it could be taken into account depending how far one went down the continued fraction before {\sl termination}.
The technique was made fully LDA-self-consistent within the tight-binding linear muffin-tin orbitals (TB-LMTO) 
approach \cite{atis} and several applications have been carried out to include short-ranged order in different alloy systems \cite{durga}. Recently Leath and co-workers have developed an itinerant CPA (ICPA) based on the
augmented space theorem \cite{glc}, which also maintains both analyticity and translational symmetry and takes into account effect of the nearest neighbour environment of a site in an alloy. The technique has been very successfully applied to the phonon
problem in alloys where there were large force constant disorders. This method  matches well with the ASR approach to the same alloys \cite{alam} and there is now a concerted effort to apply it to
electronic problems based on both the TB-KKR and the TB-LMTO methods. A very different and rather striking
approach has been developed by Rowlands \etal \cite{row2} (the non-local CPA or NL-CPA) using the idea of {\sl coarse graining}
in reciprocal space originally proposed by Jarell and Krishnamurthy \cite{jar}.  The NL-CPA with SRO  has been applied earlier by Rowlands \etal \cite{row} and 
is on the verge of being made 
 fully DFT self-consistent within the KKR. The authors report an unpublished report on it \cite{row4}.
In this communication we report a fully DFT self-consistent ASR based on the TB-LMTO with SRO incorporated.
We have applied it to the case of 50-50 CuZn alloys, so as to have a comparison with earlier attempts using different techniques.

\section{The generalized augmented space theorem}

The generalized augmented space theorem has been described in detail by Mookerjee and Prasad \cite{mp}. Let us briefly introduce those essential ideas which are necessary to make this communicated reasonably self-contained. 

For a substitutionally binary disordered alloy $A_xB_y$ on a lattice we can introduce a set of random {\sl occupation variables} $\{n_R\}$  associated with the lattice sites labelled by $R$, which take the values 0 or 1 depending upon whether the site $R$ is occupied by a A or a B type of atom. The Hamiltonian and hence the Green function are both functions of this set of random variables. The configuration averaged  Green operator  is then :

\[
 \ll G(z,\{n_R\})\gg \ =\ \int\ldots\int\ \prod_R dn_R G(z,\{n_R\})\ P(\{n_R\}) 
\]

To start with, let us assume that short-ranged order extends up to nearest neighbours only.
Let us take for an example the nearest neighbour cluster of nine atoms on a body centered cubic lattice (see Fig. \ref{figa}) centered on the site labelled by $i=0$.
Let us concentrate on the central atom (labelled 0). The occupation variables
 associated with its eight neighbours are correlated with $n_1$, but not with one another. Further none of the other occupation variables associated with more distant sites are
correlated with $n_1$.  If we label any other site $k$ as 0, that is carry out a lattice
translation from the site $i$ to $k$, its environment is
identical to the earlier situation.  Therefore the subsequent procedure is translationally symmetric, as it should be, if the SRO is itself homogeneous. We may then write :

\[
 P(n_0,n_2,\ldots n_k\ldots)  
  = P(n_0)\ \prod_{j=1}^8\ P(n_j\vert n_1)\ \prod_{k>8}\ P(n_k)
\]

The generalized Augmented Space Theorem then associates with the random variables $\{n_k\}$
corresponding operators $\{\widetilde{\bf M}_k\}$ in their configuration space. The construction of the representations of these operators has been discussed in detail in the paper by Mookerjee and Prasad \cite{mp}. Here we shall quote only the relevant results necessary to proceed further.

We shall characterize the SRO by a Warren-Cowley parameter $\alpha$. In terms of this the probability densities are given by :
\n for k=0 and k$>$8. 
\[ P(n_k) = x\delta (n_k-1)+y\delta (n_k), \quad x+y =1 \]
\n While for j=1,2,$\ldots$ 8 :
\begin{eqnarray}
P(n_j\vert n_0=1) &=& (x+\alpha y)\delta (n_j-1)+(1-\alpha)y \delta(n_j) \nonumber\\
P(n_j\vert n_0=0) &=& (1-\alpha)x\delta(n_j-1)+(y+\alpha x)\delta(n_j) \nonumber\\
\end{eqnarray}

In terms of the concentrations $x$ and $y$ and SRO parameter $\alpha$ these operators have 
 representations :
\begin{eqnarray}
{\mathbf M}_k = \left( \begin{array}{cc}
      x & \sqrt{xy}\\
    \sqrt{xy} & y \\
	\end{array} \right),\quad k=0,\ \mathrm{or}\  k > 8\phantom{XXXXX} \nonumber\\
{\mathbf M}^1_j = \left( \begin{array}{cc}
          x+\alpha y & \sqrt{(1-\alpha)y(x+\alpha y)} \\
           \sqrt{(1-\alpha)y(x+\alpha y)}& (1-\alpha)y
	\end{array}\right), \nonumber\\
{\mathbf M}^0_j = \left( \begin{array}{cc}
          (1-\alpha) x & \sqrt{(1-\alpha)x(y+\alpha x)} \\
           \sqrt{(1-\alpha)x(y+\alpha x)}& y+\alpha x
	\end{array}\right),\nonumber\\
 \quad j=1,2\ldots 8\phantom{XX} \\
\mbox{ The projection operators at the site labelled 0 are :}\nonumber \\
{\bf P}^1_0 = \left( \begin{array}{cc}
      x & \sqrt{xy}\\
    \sqrt{xy} & y \\
	\end{array} \right),\phantom{-XXXXXXXX} \nonumber\\
{\bf P}^0_0 = \left( \begin{array}{cc}
      y & -\sqrt{xy}\\
    -\sqrt{xy} & x \\
	\end{array} \right),\phantom{XXXXXXX} 
\end{eqnarray}

In the full augmented space, which is the product space of all the individual configuration
space of the different occupation variables, the operators which replace the occupation variables are :

\begin{eqnarray}
\widetilde{\mathbf M}_0  =&  M_1\otimes I\otimes I\otimes \ldots \quad\quad\phantom{XXXXX} \nonumber \\
\widetilde{\mathbf M}_j  =&\sum_{k=0}^1 P^k_1\otimes M_2^k \otimes I\otimes \ldots 
\quad \quad \mbox{\rm j=1,2,$\ldots$ 8} \nonumber \\
\widetilde{\mathbf M}_{k}   = & I\otimes I \otimes \ldots M_k\otimes I\otimes \ldots 
 \quad\quad \mbox{\rm k $>$ 8} \nonumber \\
\end{eqnarray}

In operator form the operators labelled by $0$ and $k\ >\ 8$ have the same form as in Equation (\ref{asr1}). The operators labelled by $j$ have the form :

\begin{widetext}
\begin{eqnarray}
\widetilde{\mathbf M}_j & = & x\ {\cal I}\otimes{\cal I}\ldots\ +\ b(x)\
{\cal P}^\downarrow_j\otimes{\cal I}\ldots +\ b'(x)\ {\cal P}^\downarrow_0\otimes {\cal I}\ldots\ +\ f^{\prime\prime}(x)\ {\cal T}^{\uparrow\downarrow}_0\otimes{\cal I}\ldots\ +\ b^{\prime\prime}(x)\ {\cal P}^\downarrow_0\otimes{\cal P}^\downarrow_j\otimes\ldots\nonumber\\
& & +\ f^\prime(x)\ {\cal T}^{\uparrow\downarrow}_j\otimes{\cal I}\ldots\ +\
d^{\prime\prime}(x)\ {\cal P}^\downarrow_0\otimes{\cal T}^{\uparrow\downarrow}_j\ldots\ +\ d(x)\ {\cal T}^{\uparrow\downarrow}_0\otimes{\cal P}^\downarrow_j\ldots\ +\ f^{\prime\prime\prime}(x)\ {\cal T}^{\uparrow\downarrow}_0\otimes{\cal T}^{\uparrow\downarrow}_j\ldots \nonumber\\
\label{asr2}
\end{eqnarray}
\n where,
\begin{eqnarray*}
& b'(x) = \alpha (y-x) \qquad
 b^{\prime\prime}(x) = -2\alpha (y-x)\qquad
 F^\prime(x) = (x\beta_1+y\beta_2) \qquad
 F^{\prime\prime}(x) = \alpha\sqrt{xy}  \\
& F^{\prime\prime\prime}(x) = (\beta_1-\beta_2)\sqrt{xy};\qquad
 D(x) = -2\alpha \sqrt{xy}; \qquad
 D^{\prime\prime}(x) = (y-x)(\beta_1-\beta_2)\\ 
& \beta_1 = \sqrt{(1-\alpha)y(x+\alpha y)}\ ;\
\beta_2 = \sqrt{(1-\alpha)x(y+\alpha x)}\\
\end{eqnarray*}
\end{widetext}

\begin{figure*}
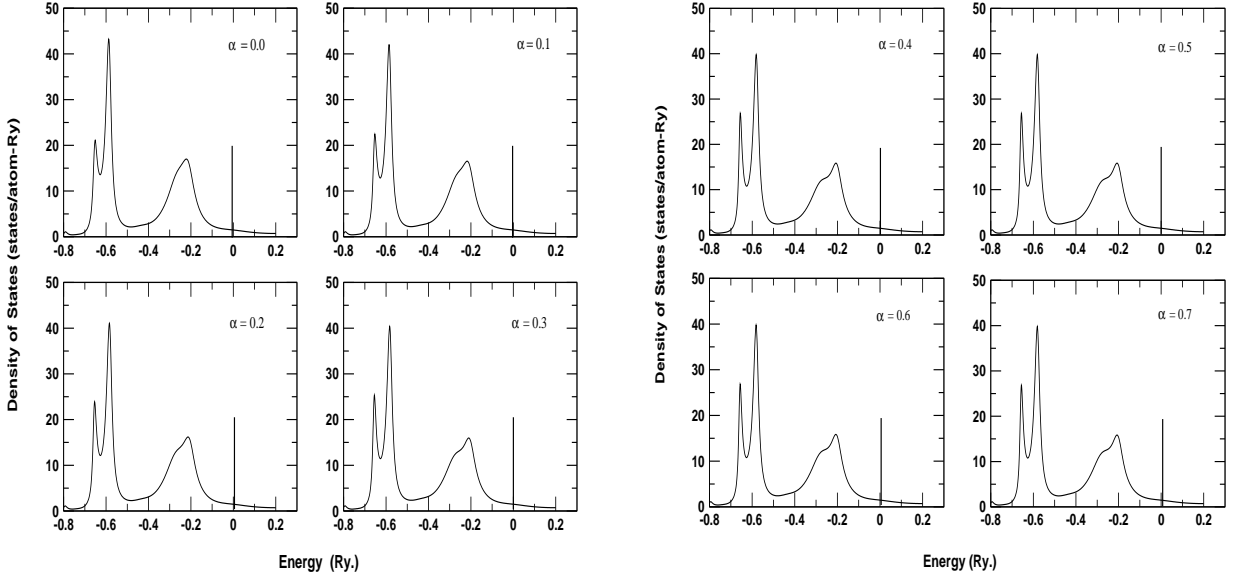

\centering
\vskip 0.8cm
\includegraphics[width=3in,height=3in]{fig9.eps}\hskip 1cm 
\includegraphics[width=3in,height=3in]{fig10.eps}
\caption{Density of states for 50-50 CuZn with increasing positive $\alpha$ which describes
increasing clustering tendency}
\label{fig10}
\vskip 0.5in
\end{figure*}

\begin{figure*}
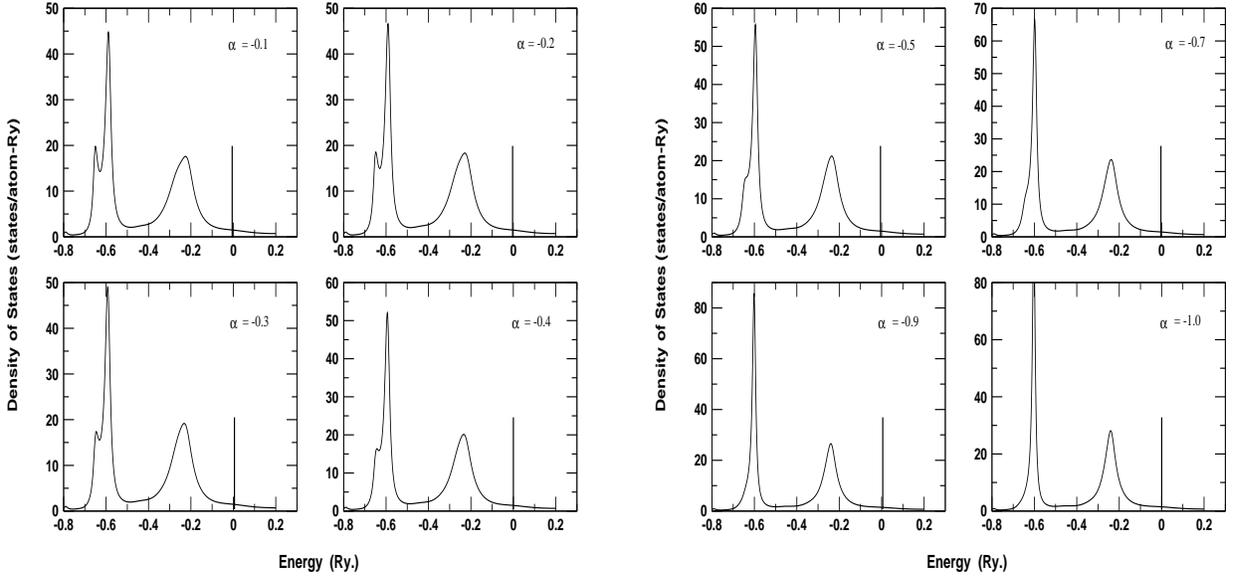

\centering
\vskip 0.8cm
\includegraphics[width=3in,height=3in]{fig11.eps}\hskip 1cm 
\includegraphics[width=3in,height=3in]{fig12.eps}
\caption{Density of states for 50-50 CuZn with increasing negative $\alpha$ which describes
increasing ordering tendency}
\label{fig11}
\end{figure*}

We now follow the augmented space theorem and replace all the occupation variables $\{n_R\}$ by their corresponding operators.
 
It is easy to check that in the absence of short-ranged order the functions
$b'(x),\ b^{\prime\prime}(x),\ f^{\prime\prime}(x),\ f^{\prime\prime\prime}(x),\ d(x)$ and $ d^\prime(x)$ all vanish, $\beta_1 =\beta_2=\sqrt{xy}$ and $f^\prime(x)=f(x)$.  We also note that the choice of the {\sl central} site labelled `0' is immaterial. If we translate this site to any other and apply the lattice translation to
all the sites,  hence the Hamiltonian in the full augmented space, remains unchanged. This formulation of short ranged order also possesses lattice translational symmetry, provided the short-ranged order is homogeneous in space.

\section{Effect of SRO on the Density of States}

We have carried out the TB-LMTO-ASR  calculations on CuZn with a lattice constant of 2.85 \AA. The Cu and Zn potentials are self-consistently
obtained via the LDA self-consistency loop. All reciprocal space integrals
are done by using the generalized tetrahedron integration for disordered systems introduced by us earlier \cite{kspace}. 

 To discuss the effect of SRO, leading, on one hand, to ordering ($\alpha < $ 0)
 and  segregation on the other ($\alpha >$ 0), let us first look at Figs. \ref{fig1} and \ref{fig3}.  The complex band structure shown in Fig. \ref{fig1} shows
that the system is a {\sl split band} alloy. The positions of the $d$-bands of
Cu and Zn are well separated in energy. This implies that the ``electrons travel
more easily between Cu or between Zn sites than between unlike ones" \cite{row}. So when the alloy orders and unlike sites sit next to each other, the overlap
integral between the like sites decrease. This leads to a narrowing of the bands associated with Cu and Zn. A comparison between the leftmost and central panels of Fig. \ref{fig3} shows that the bandwidths in the latter are almost half of the former. This should be the main effect of ordering setting in.  
 On the other hand, when the alloy is completely disordered, the bands gets widened by disorder scattering and the sharp structures in the density of states are smoothened. 

Fig (\ref{fig10}) shows the density of states with increasing positive $\alpha$. Positive $\alpha$ indicates a clustering tendency. Comparing with Fig. (\ref{fig3}) we note that as clustering tendency increases the density of states begins
to show the structures seen in the pure metals in both the split bands. For large positive $\alpha$ there is still residual long-ranged disorder. This causes smoothening of the bands with respect to the pure materials. For these large,
positive $\alpha$s, we notice the development of the structure around -0.3Ry.

Fig. (\ref{fig11}) shows the density of states with increasing negative $\alpha$ which indicates increasing ordering tendency. On the bcc lattice at 50-50 composition we expect this ordering to favour a B2 structure. With increasing ordering tendency, both the split bands narrow and lose structure. The feature around
-0.3 Ry. disappears. This band narrowing and suppression of the feature around
-0.3 Ry. is clearly seen in the ordered B2 alloy shown in Fig. (\ref{fig3}). 

\begin{figure}
\centering
\vskip 0.8cm
\includegraphics[width=2.5in,height=2.5in]{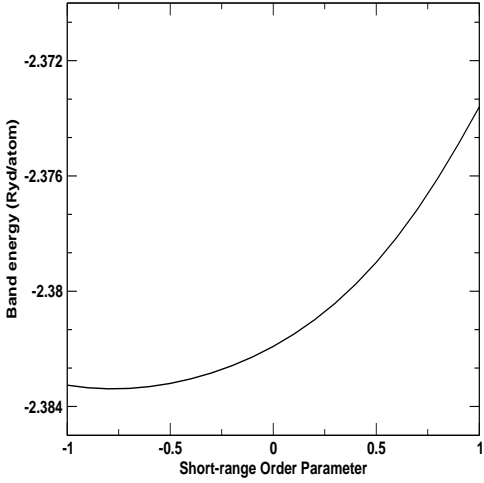}
\caption{The Band energy (from which the contribution of the core electrons
have been subtracted) as a function of the nearest-neighbour Warren-Cowley SRO parameter}
\label{stab}
\end{figure}

Our analysis is closely similar to that of Rowlands \etal \cite{row}. However,
there are quantitative differences as both the basic technique (KKR and TB-LMTO) and the approximations are quantitatively different.
 Finally, in Fig. \ref{stab} we show the band energy as a function of the
nearest neighbour Warren-Cowley parameter. The minimum occurs at the ordering end, as expected. Experimentally the alloy does show a tendency to order at lower temperatures.

\section{Optical properties of CuZn alloys}
In an earlier work \cite{tm} we had developed a methodology for the calculation of configuration averaged optical conductivity of a disordered alloy based on the augmented space formalism. Here we shall present the salient features required for the calculation of optical response functions in the 50-50 CuZn alloy. 

In linear response theory, at zero temperature, the generalized susceptibility of a disordered alloy is given by the Kubo formula : 

\[ \langle \j^\mu(t)\rangle = \int_{-\infty}^\infty \chi^{\mu\nu}(t-t')\ A^\nu(t') \]

\n where, $A^\nu(t)$ is the vector potential, and 

\[ \chi^{\mu\nu}(t-t') = (i/\hbar)\ \Theta(t-t')\ \langle[\j^\mu(t) \j^\nu(t')]\rangle \]

\n {\bf j}$^\mu$ is the current operator and $\Theta$ is the Heaviside step function.
If the underlying lattice has cubic symmetry, $\chi^{\mu\nu}$ = $\chi\ \delta_{\mu\nu}$.
The fluctuation dissipation theorem then relates the imaginary part of the Laplace
transform of the generalized  susceptibility to the Laplace transform of a correlation
function :

\[
\chi^{\prime\prime}(\omega) = (1/2\hbar) \left(1-\exp\{-\beta\hbar\omega\}\right)S(\omega) \]
\n where,

\begin{equation}
 S(\omega) = \int_0^\infty\ dt \ \exp\{i(\omega + i\delta)t\}\ \Tr \left(\rule{0mm}{4mm} \j^\mu(t)\ \j^\mu(0)\ \right)
\end{equation}

Since the response function is independent of the direction label $\mu$ for cubic symmetry, in the
following we shall drop this symbol. In case of other symmetries we have to generalize our results
for different directions.
Our goal will be, given a quantum ``Hamiltonian" {\bf H} to obtain the
correlation function : 
$S(t) = \langle \phi\vert \mathbf{j}(t)\mathbf{j}(0)\vert\phi\rangle$

We shall now determine the correlation directly via the recursion method as described by Viswanath and M\"uller\cite{gb}. 

For a disordered binary
alloy,  $S(t)= S[\bar{\mathbf H}(\{n_R\})]$. The augmented space theorem then
states that :

\begin{eqnarray*}
 \ll S(t)\gg   =  \ll \langle \phi\vert {\mathbf j}(t){\mathbf j}(0)\vert\phi\rangle \gg\phantom{XXXXXXXX}\\
  = \langle \phi\otimes \{\emptyset\}\vert \tilde{\mathbf j}(t)\tilde{\mathbf j}(0)\vert\phi\otimes\{\emptyset\}\rangle = S[\widetilde{\mathbf H}(\{\widetilde{\mathbf M}^R\})]
\end{eqnarray*} 

\n where the augmented space Hamiltonian and the current operators are constructed by replacing every random variable $n_R$ by the corresponding operator $\widetilde{\bf M}^R$.

\n The recursion may now be modified step by step in the full augmented space :

\[ \langle \psi(t)\vert\ =\ \langle\phi\otimes\{\emptyset\}\vert\ \tilde{\mathbf{j}}(t) \]

\n The time evolution of this {\sl bra} is governed by the Schr\"odinger equation

\begin{equation}
 -i\ \frac{d}{dt}\left\{\rule{0mm}{4mm} \langle\psi(t)\vert\right\}\ =\ \langle\psi(t)\vert\widetilde{\mathbf {H}} 
\label{eq2a}
\end{equation}

As before, we shall generate the orthogonal basis $\{\langle f_k\vert\}$ for representation of equation (\ref{eq2a}) :

\begin{enumerate}
\item[(i)]We begin with initial conditions : \[ \langle f_{-1}\vert = 0\quad ;\quad \langle f_0\vert = \langle \phi\otimes\{\emptyset\}\vert\tilde{\bf j}(0)\]
\item[(ii)] The new basis members are generated by a three term recurrence
relationship :
\[ \langle f_{k+1}\vert\ = \ \langle f_k\vert \widetilde{\mathbf H}\ - \ \langle f_k\vert\ \tilde{\alpha}_k \ -\ \langle f_{k-1}\vert\ \tilde{\beta}_k^2\quad\quad \mathrm {k=0,1,2}\ldots\]

\n where, \begin{equation}
 \tilde{\alpha}_k = \frac{\langle f_k \vert\widetilde{\bf H}\vert f_k\rangle}{\langle f_k \vert f_k\rangle}\quad\quad \tilde{\beta}_k^2= \frac{\langle f_k\vert f_k\rangle}{\langle f_{k-1}\vert f_{k-1}\rangle } 
\label{opt1}
\end{equation}
\end{enumerate}

\n We now expand the bra $\langle \psi(t)\vert $ in this orthogonal basis :

\[
 \langle \psi(t)\otimes\{\emptyset\}\vert \ =\ \sum_{k=0}^\infty\ \langle f_k\vert\ \widetilde{D}_k(t) 
\]

\n Continuing exactly as in the last section we get,
\begin{equation}
 \widetilde{D}_0(t) = \langle\psi(t)\otimes \{\emptyset\}\vert f_0\rangle
 = \ll S(t)\gg
\label{opt2}
\end{equation}

\n Taking Laplace transforms and using the three term recurrence,

\begin{equation}
\tilde{d}_0(z) \ =\ \frac{i}{\displaystyle z-\tilde{\alpha}_0-\frac{\tilde{\beta}^2_1}{
\displaystyle{z-\tilde{\alpha}_1- \frac{\tilde{\beta}^2_2}{\displaystyle z-\tilde{\alpha}_2 - \ldots}}}}
\label{opt3}
\end{equation}

The configuration averaged structure function, which is the Laplace transform of the averaged correlation function can then be obtained from the above :

\begin{equation}
\ll S(\omega)\gg \ =\ \lim_{\delta\rightarrow 0}\ 2\ \Re e\ \tilde{d}_0(\omega+i\delta) 
\label{eq6a}\end{equation}

The imaginary part of the dielectric function is related to this correlation
function through :

\[ \epsilon_2(\omega)\ =\ \frac{\ll S(\omega)\gg}{\omega)} \]

The real part of the dielectric function is related to the imaginary part through a Kramers-Kr\"onig relationship.  
The equations (\ref{opt1})-(\ref{eq6a}) will form the basis of our calculation of the configuration averaged correlation function.

\n Next we look at the expression for the current operator in the TB-LMTO basis :

\[
\vert\chi_R\rangle=\vert\phi_R\rangle +\sum_{R^{\prime}}h_{RR\prime}\vert\dot{\phi}_{R^{\prime}}\rangle
\]

\n The dot refers to derivative with respect to energy.
In this basis, the matrix elements of the current operator can be written as

\[
 J^\mu_{RR^{\prime}}=\langle\chi_{R^{\prime}}\vert {\mathbf j}^\mu\vert\chi_{R}\rangle \]
where,
\begin{eqnarray}
 {J^\mu_{RR^{\prime}}}=  e\left[V^{(1),\mu}_{RR^{\prime}}\ \delta_{RR^{\prime}}+\sum_{R^{\prime\prime}}V^{(2),\mu}_{RR^{\prime\prime}}\ h_{R^{\prime\prime}R^{\prime}}+ \ldots\right.\phantom{XXXX}\nonumber\\
\left. \sum_{R^{\prime\prime}}h_{RR^{\prime\prime}}\ V^{(3),\mu}_{R^{\prime\prime}R{\prime}}+\sum_{R^{\prime\prime}}\sum_{R^{\prime\prime\prime}}h_{RR{\prime\prime\prime}}\ V^{(4),\mu}_{R^{\prime\prime\prime}R{\prime\prime}}\ h_{R^{\prime\prime}R^{\prime}}\right]\nonumber\\
\end{eqnarray}

\n with,
\begin{eqnarray*}
 V^{(1),\mu}_{RR^{\prime}}=\langle\phi_{R^{\prime}}\vert {\mathbf v}^\mu\vert\phi_{R}\rangle \ ;\ 
 V^{(2),\mu}_{RR^{\prime}}=\langle\dot{\phi}_{R^{\prime}}\vert {\mathbf v}^\mu\vert\phi_{R}\rangle \\ 
 V^{(3),\mu}_{RR^{\prime}}=\langle\phi_{R^{\prime}}\vert {\mathbf v}^\mu\vert\dot{\phi}_{R}\rangle \ ;\ 
 V^{(4),\mu}_{RR^{\prime}}=\langle\dot{\phi}_{R^{\prime}}\vert{\mathbf v}^\mu\vert\dot{\phi}_{R}\rangle
\end{eqnarray*}

The technique for calculating these matrix elements has been described in detailed by Hobbs \etal \cite{hobbs}. We have also used this technique in our earlier
paper \cite{km1} and we shall use it here as well. The readers are referred to these two papers for details.

Ideally the next step would be  to calculate $J^\mu_{AA},J^\mu_{BB},J^\mu_{AB},J^\mu_{BA}$ as the current terms between two sites when they are occupied by atom pairs
AA, BB, AB and BA embedded in the disordered medium. A simpler first step would
be to obtain these current terms from the pure A and B and from the ordered AB alloy. In general, the current operator can be written as :

\be {\mathbf j}^\mu = \sum_R J^\mu(0)\ {\bf P}_R + \sum_R\sum_{R'}\ J^\mu(\chi)\ {\bf T}_{RR'}\ee

\n where $\chi = R-R'$.

In a disordered alloy, the representations of the current operators are random :

\begin{eqnarray*}
J^\mu(0)&=&J^\mu_{AA}(0)\ n_R+J^\mu_{BB}(0)\ (1-n_R) \qquad  \mathrm {and}\\
J^\mu(\chi)&=&J^\mu _{AA}(\chi)\ n_R\ n_{R'} +J^\mu _{BB}(\chi)\ (1-n_R)\ (1-n_{R'}) + \\
 &&J^\mu _{AB}(\chi)\ n_R\ (1-n_{R'})+J\mu_{BA}(\chi)\ (1-n_R) n_{R'}
\end{eqnarray*}

Where $R-R'\ =\ \chi$.
Using the augmented space theorem,  we replace the random variable $n_R$ by an operator $M^R$ in the expressions for $J^\mu(0)$ and $J^\mu(\chi)$ we get 
the current operator in the augmented space~:
\begin{figure*}
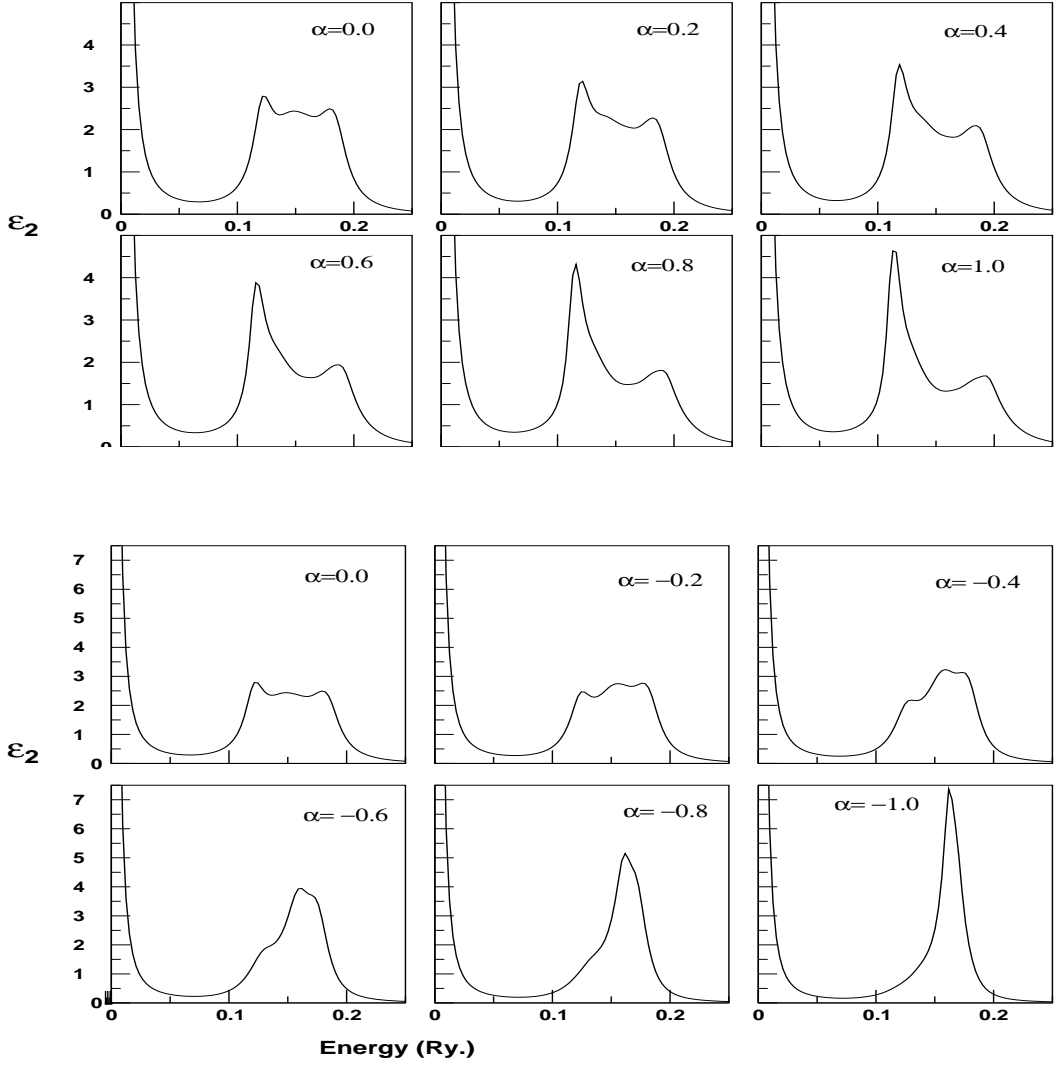

\centering
\vskip 0.7cm
\includegraphics[width=5.5in,height=2.7in]{fig14.eps}
\vskip 0.4cm
\includegraphics[width=5.5in,height=2.7in]{fig15.eps}
\caption{(Top panel)Imaginary part of the dielectric function as a function of energy and its variation with the nearest neighbour Warren-Cowley SRO parameter,$\alpha\ >\ $ 0 increasing,  indicating
segregating tendency. (Bottom panel)
The same with $\alpha\ <\ $ 0 increasing, indicating ordering tendency} 
\label{epsP}
\end{figure*}

\begin{widetext}
\begin{eqnarray}
\tilde{\mathbf j}^\mu = & \sum_R \ \left\{\rule{0mm}{4mm} J^\mu_{BB}(0) {\cal P}_R
\otimes \widetilde{\bf I}\ldots + J_1^\mu(0) {\cal P}_R\otimes\widetilde{\bf M}_R \right\}+ 
\sum_R\sum_{R'} \left\{\rule{0mm}{4mm} J^\mu_{BB}(\chi)\ {\cal T}_{RR'}\otimes\widetilde{\mathbf I} +
J_1^\mu(\chi)\ {\cal T}_{RR'}\otimes\left(\widetilde{\bf M}_R + \widetilde{\bf M}_{R'}\right) + \right.\nonumber\\
 & \left. J_2^\mu(\chi)\ {\cal T}_{RR'}\otimes\widetilde{\bf M}_R\otimes\widetilde{\bf M}_{R'} \rule{0mm}{4mm}\right\}\phantom{XXXXXXXX}
\end{eqnarray}

\begin{eqnarray*}
J^\mu_1(0)=[J_{AA}^\mu(0)-J_{BB}^\mu(0)];\quad
J^\mu_2(0)=[J_{AA}^\mu(0)-J_{BB}^\mu(0)] \\
J^\mu_1(\chi)=[J_{AA}^\mu(\chi)-J_{BB}^\mu(\chi)];\quad
J^\mu_2(\chi)=[J_{AA}^\mu(\chi)-J_{BB}^\mu(\chi)]\\ 
J_3^\mu(\chi)=J_{AA}^\mu(\chi)-J_{AB}^\mu(\chi)-J_{BA}^\mu(\chi)+J_{BB}^\mu(\chi) 
\end{eqnarray*}
\end{widetext}

\begin{figure*}
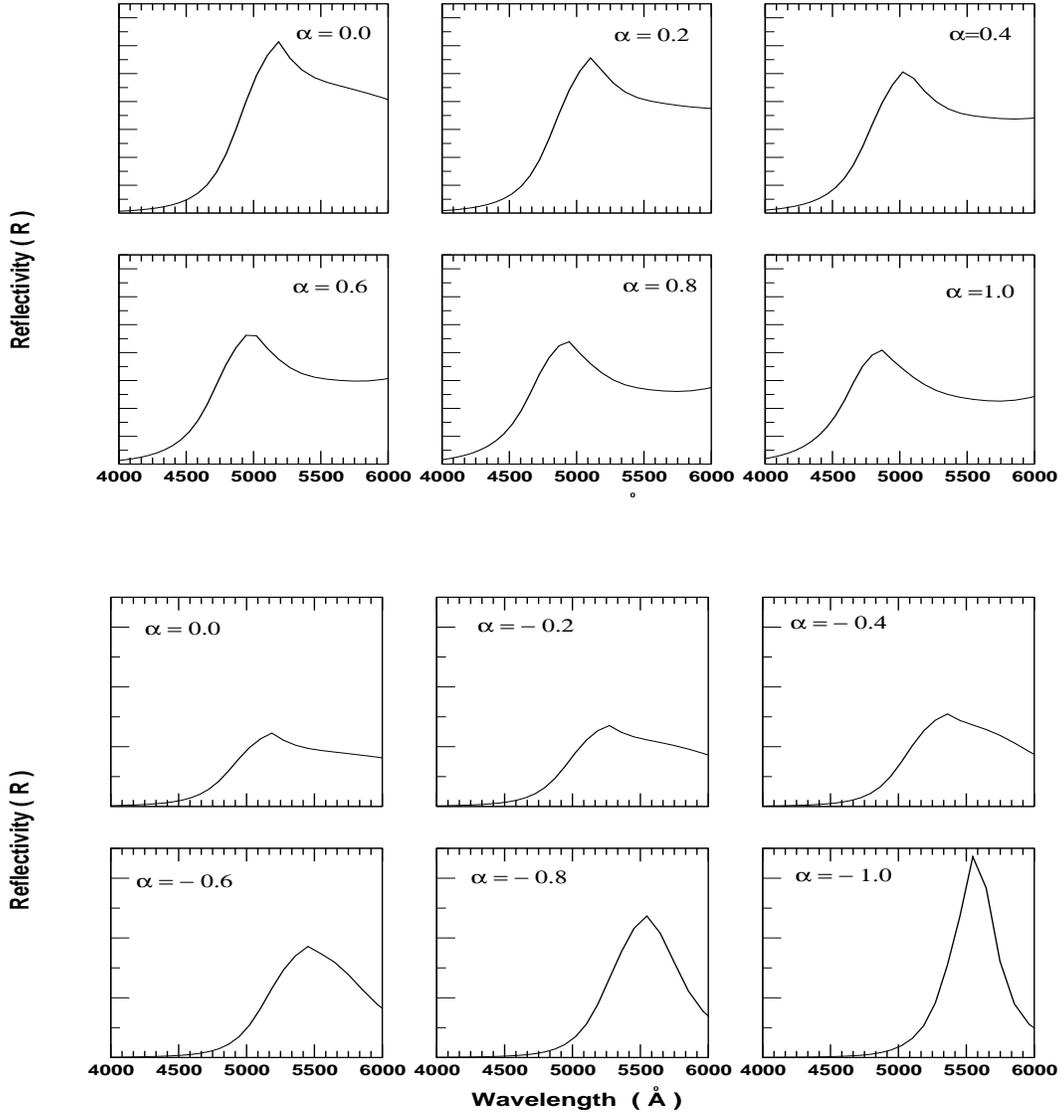

\centering
\includegraphics[width=5.5in,height=2.7in]{fig16.eps}
\vskip 1cm
\includegraphics[width=5.5in,height=2.7in]{fig17.eps}
\caption{(Top panel)Reflectivity as a function of the wavelength and its variation with the nearest neighbour Warren-Cowley parameter $\alpha\ >\ $ 0 increasing indicating segregating tendency. (Bottom panel)
The same with $\alpha\ <\ $ 0 indicating ordering tendency} 
\label{ref}
\end{figure*}

The relevant operators are now chosen from  Equations (\ref{asr1}) and (\ref{asr2}) and the current operator in full augmented space is set up from them.  
This augmented current operator is used to construct the starting state of the recursion as described in the last section.

Since the real part of the dielectric function $\epsilon_1(\omega)$ can be
now obtained from  the  imaginary part $\epsilon_2(\omega)$ using a Kramers Kr\"onig relationship, we know the full complex dielectric function. All optical response functions may now be derived from these. If we assume the orientation of the crystal surface to be parallel to the optic axis, the reflectivity $R(\omega)$ follows directly from Fresnel's formula :

\[ R(\omega)\ =\ \left | \frac{\sqrt{\epsilon(\omega)}-1}{\sqrt{\epsilon(\omega)}+1}\right |^2 \]

 Fig. \ref{epsP} (top panel) shows the imaginary part of the dielectric function varying with frequency showing increasing segregating tendency. If we examine the density of states for both the ordered and
disordered alloys we note that it is only around or just above $\hbar\omega \simeq$ 0.1 Ry that the transitions from the $d$-bands of Cu begin to contribute
to $\epsilon_2(\omega)$. Below this energy, the behaviour is Drude like. This is clearly seen in the panels of the figure. As $\alpha$ increases and the alloy
tends to segregate, since the weight of the structure in the density of states 
nearer to the Fermi-energy, increases, this contribution leads to the increasing
weight of the structure near 0.1 Ry. 

Fig. \ref{epsP} (bottom panel) shows the variation of $\epsilon_2(\omega)$ with the nearest neighbour Warren-Cowley SRO parameter showing increased ordering
tendency. The narrowing of the band with ordering and the increase of weightage to the structure away from the Fermi energy is clearly reflected in the dielectric function. It is also obvious from the figures that
the imaginary part of the dielectric function is more sensitive to the variation
of short-ranged order than the density of states.

Color changes in CuZn brasses have been observed both with changing composition,
hence different phase structures, and on heating \cite{mul}. In particular as these alloys are coloured and the changes can be observed visually. A study of
the changes in  spectral reflectivity due to ordering or segregation  would be interesting. The {\sl colour} of these alloys should be due to the same physical mechanism of internal photoelectric excitations, as proposed for coloured metals like Cu and Au. Of course, we should take care of plasma effects and other
phenomena that together and in a combined way give rise to {\sl colour}. However, it would be, as an initial study, examine the effect of SRO on the spectral
reflectivity of the alloy.  This is sown in Fig. \ref{ref}.

The first thing we note that for the completely disordered alloy the reflectivity has a maximum around $\lambda = $5200~\AA\  which is not very different from
that of Cu at low temperatures and is in the Yellow-Green region. Below this the reflectivity sharply drops (the reflectivity edge) and light of lower wavelengths is not reflected. With increasing segregation tendency this maximum drops almost linearly with $\alpha$ to around 4950 \AA\ which is towards the green range. On the other hand with increasing ordering tendency the maximum moves out towards
5600\AA\ which moves towards the yellow from the yellow-green wavelengths. We would expect then a shift from a yellow-green colour in the disordered to a more
yellow region in the ordered material. We should note that these calculations are all done at 0K. In an actual experiment the disorder-order transition takes
place with varying temperature and the temperature effects on the electronic
structure must be taken into account, as well as effect of plasma oscillations.
We shall leave this for a later communication.

\end{document}